\input epsf
\documentstyle[12pt]{article}
\newlength{\dinwidth}
\newlength{\dinmargin}
\def\Journal#1#2#3#4{{#1} {\bf #2}, #3 (#4)}

\def\NPB{{\em Nucl. Phys.} B}
\def\PLB{{\em Phys. Lett.}  B}
\def\PRL{\em Phys. Rev. Lett.}
\def\PRD{{\em Phys. Rev.} D}
\def\ZPC{{\em Z. Phys.} C}
\def\PTP{\em Prog. Theor. Phys.}

\newcommand{\ra}{\rightarrow}
\newcommand{\BGAMAXS}{B \ra X _{s} + \gamma}
\newcommand{\BGAMAXD}{B \ra X _{d} + \gamma}
\newcommand{\BbarGAMAXS}{\overline{B} \ra \overline{X_s} + \gamma}
\newcommand{\BbarGAMAXD}{\overline{B} \ra \overline{X_d} + \gamma}
\newcommand{\BBGAMAXS}{{\cal B}(B \ra  X _{s} + \gamma)}
\newcommand{\BBGAMAXD}{{\cal B}(B \ra  X _{d} + \gamma)}
\newcommand{\BBbarGAMAXS}{{\cal B}(\overline{B} \ra  \overline{X_s} + \gamma)}
\newcommand{\BBbarGAMAXD}{{\cal B}(\overline{B} \ra  \overline{X_d} + \gamma)}

\newcommand{\BGAMAKSTAR}{B \ra  K^{\star} + \gamma}

\newcommand{\BRAV}{\langle {\cal B}(B \ra  X_d + \gamma)\rangle}

\newcommand{\GbarGAMAXS}{\Gamma (\overline{B} \ra  \overline{X_s} + \gamma)}

\newcommand{\absvcb}{\vert V_{cb}\vert}

\newcommand{\absvts}{\vert V_{ts}\vert}
\newcommand{\absvtb}{\vert V_{tb}\vert}

\newcommand{\ba}{\begin{array}}
\newcommand{\ea}{\end{array}}
\newcommand{\be}{\begin{equation}}
\newcommand{\ee}{\end{equation}}
\newcommand{\bea}{\begin{eqnarray}}
\newcommand{\eea}{\end{eqnarray}}

\def\l{\lambda}

\def\to{\rightarrow}
\def\mb{m_b}

\def\as{\alpha _s}

\begin{document}
\thispagestyle{empty}
\addtocounter{page}{-1}
\begin{flushright}
DESY 97-255\\
BUTP-98/08\\
March 1998
\end{flushright}
\vspace*{1.8cm}
\begin{center}
{\large\bf Inclusive Decay Rate for $B \to X_d +
\gamma$ in
Next-to-Leading Logarithmic Order and CP Asymmetry in the Standard Model
\footnote{Work partially supported by Schweizerischer Nationalfonds.}}
\end{center}
\vspace*{1.0cm}
\centerline{\large\bf A.~Ali$^a$, H.~Asatrian$^b$ and C. Greub$^c$}
\vspace*{0.5cm}
\centerline{\large\bf {\rm ${}^a$}Deutsches Elektronen-Synchrotron
DESY,  Hamburg, Germany}
\centerline{\large \bf {\rm ${}^b$}Yerevan Physics Institute,
   Alikhanyan Br., 375036-Yerevan, Armenia}
\centerline{\large \bf {\rm ${}^c$} Inst. f. Theor. Physik, Univ.
Bern, Bern, Switzerland}
\vspace*{1.0cm}
\centerline{\Large\bf Abstract}
\vspace*{1cm}
We compute the decay rate for the Cabibbo-Kobayashi-Maskawa (CKM)-suppressed
electromagnetic penguin decay $\overline{B} \to
 \overline{X_d} + \gamma$ (and its charge conjugate) in the next-to-leading
order in
QCD, including leading power corrections in $1/m_b^2$ and $1/m_c^2$ in the
standard model.
The average branching ratio $\BRAV$ of the decay $\BGAMAXD$ and 
its charge conjugate $\BbarGAMAXD$  
is estimated to be in the range $6.0 \times 10^{-6} \leq
\BRAV 
\leq 2.6 \times 10^{-5}$, obtained by varying
the CKM-Wolfenstein parameters $\rho$ and $\eta$ in the range
$-0.1 \leq \rho \leq 0.4$ and $0.2 \leq \eta \leq 0.46$ and taking into
account other parametric dependence. In the NLL approximation and in the
stated range of the CKM parameters, we find the ratio 
$R(d\gamma/s\gamma) \equiv
\langle {\cal B}(B \to X_d\gamma)\rangle/\langle {\cal B}(B \to X_s\gamma)$
to lie in the range $0.017 \leq R(d\gamma/s\gamma) \leq 0.074$.
Theoretical uncertainties in this ratio
are estimated and found to be small. Hence, this ratio is well suited to
provide independent constraints on the CKM parameters.
 The CP-asymmetry in the decay rates,
defined as $a_{CP}(B \to X_d\gamma)\equiv (\Gamma(B \to X_d\gamma) -
\Gamma(\overline{B} \to \overline{X_d} \gamma))/(\Gamma(B \to X_d\gamma) +
\Gamma(\overline{B} \to \overline{X_d} \gamma))$,  is found to be in the 
range $(7 - 35)\%$. Both the decay rates and CP asymmetry are measurable in
forthcoming experiments at $B$ factories and possibly at HERA-B.
 
\newpage
{\bf 1. Introduction}

Electromagnetic penguins were first measured
by the CLEO collaboration through the exclusive
decay $\BGAMAKSTAR$ \cite{CLEOrare1},
followed by the measurement of the inclusive
decay $\BGAMAXS$ \cite{CLEOrare2}.
The present CLEO measurements can be summarized as \cite{Tomasz97}:
\begin{eqnarray}
\label{penguinexp}
\langle {\cal B}(\BGAMAXS)\rangle &=& (2.32\pm 0.57\pm 0.35)\times 10^{-4}, 
\nonumber\\
\langle {\cal B}(\BGAMAKSTAR)\rangle &=& (4.2\pm 0.8 \pm 0.6)\times 10^{-5}.
\end{eqnarray}
Very recently, the
inclusive radiative decay has also been reported by the ALEPH collaboration
with a (preliminary) branching ratio \cite{ALEPHbsg}:
\begin{equation}
\label{alephbsg}
\langle{\cal B}(H_b \to X_s + \gamma)\rangle = (3.29 \pm 0.71 \pm 
0.68)\times 10^{-4}.
\end{equation}
The quantity $\langle {\cal B}(\BGAMAXS) \rangle$ is the
branching ratio 
averaged over the decays $\BGAMAXS$ and its charge conjugate
$\BbarGAMAXS$.
 The  branching ratio in (\ref{alephbsg}) involves a
different weighted average of the $B$-mesons and $\Lambda_b$ baryons
produced in $Z^0$ decays (hence the symbol $H_b$ ) than the
corresponding one
given in (\ref{penguinexp}), which has been measured in the decay
$\Upsilon (4S) \to B^+ B^-, B^0 \overline{B^0}$.

 These measurements have stimulated an
impressive theoretical activity, directed at improving the
precision of the decay rates and distributions in the context of the
standard model (SM) and beyond the standard model,
in particular supersymmetry.
In the SM-context, the complete next-to-leading-logarithmic
(NLL) contributions have been painstakingly
completed \cite{ag1} - \cite{CDGG97},
and leading power corrections in $1/m_b^2$
\cite{georgi,manoharwise,FLS94} and $1/m_c^2$
\cite{Voloshinbsg97,powermc,BIR97} have also been calculated for the decay
rate in $\BGAMAXS$. This theoretical work allows to calculate the
branching ratios in the SM with an accuracy of about $\pm 9\%$, yielding
$\langle {\cal B}(\BGAMAXS ) \rangle =
(3.50 \pm 0.32) \times 10^{-4}$ and 
$\langle {\cal B}(H_b \to X_s + \gamma) \rangle =
(3.76 \pm 0.30) \times 10^{-4}$,  in reasonable agreement
with the CLEO and (preliminary) ALEPH measurements, respectively.
 The decay rates in
eqs.~(\ref{penguinexp}) and (\ref{alephbsg}) determine the
ratio of the Cabibbo-Kobayashi-Maskawa (CKM) \cite{CKM} matrix elements
$\vert V_{ts}^*V_{tb}/V_{cb}\vert$. Since $\absvcb$ and $\absvtb$ have been
directly measured, these measurements can be combined, yielding
$\absvts =0.033 \pm 0.007$ \cite{alisb97}. The central value of $\absvts$
is somewhat lower than the corresponding
value of $\absvcb$, $\absvcb=0.0393 \pm 0.0028$, but within errors
the two matrix elements are found to be approximately equal, as expected
from the CKM unitarity.

 The interest in measuring the decay rate in $\BGAMAXD$
(and its charge conjugate  $\BbarGAMAXD$) lies
in that it will determine the CKM-Wolfenstein parameters $\rho$ and $\eta$
\cite{Wolfenstein} in a theoretically reliable way.
 Likewise, this decay
will enable us to search for new physics which may manifest itself through
enhanced $bd\gamma$ and/or $bdg$ effective vertices. These
vertices are CKM-suppressed in the standard model, but new physics
contributions may not follow the CKM pattern in
flavor-changing-neutral-current transitions and hence new physics effects
may become more easily discernible in $\BGAMAXD$ (and its charge 
conjugate) than in the
corresponding CKM-allowed vertices $bs\gamma$ and $bsg$.
Closely related to this
is the question of CP-violating asymmetry in the decay rates for
$\overline{B} \to \overline{X_d} + \gamma$ and its charge conjugate
$B \to X_d + \gamma$, which 
may provide us with the first measurements of the so-called direct CP
violation in $B$ physics. With the weak phase provided dominantly
by the CKM matrix elements $V_{td}$ and $V_{ub}$ in the decay
$\BGAMAXD$, the perturbatively generated strong phases can be
calculated by taking into account the charm and up
quark loops in the electromagnetic penguins, which generate the necessary
absorptive contributions.  This calls for
an improved theoretical estimate of $\BBGAMAXD$ and 
$\BBbarGAMAXD$ (hence $a_{CP}$) in the SM.

In what follows, we shall discuss for the sake of definiteness the
decays of the $b$-quark $b \to s + \gamma (+g)$ and $b \to d + \gamma(+g)$, 
whose hadronic transcriptions are $\BbarGAMAXS$ and $\BbarGAMAXD$,
respectively, but the discussion applies for the charge conjugate decays 
$\BGAMAXS$ 
and $\BGAMAXD$ as well with obvious changes. When it is essential 
to differentiate between the decay of a
$B$ meson and its charge conjugate $\overline{B}$, we shall do so.
The branching ratio for the CKM-suppressed decay $\BbarGAMAXD$ was calculated
several years ago in partial next-to-leading order by two of us (A.A. and C.G.)
\cite{ag2}.  Subsequent to that,
the CP asymmetries in the decay rates in the leading logarithmic (LL)
approximation were calculated in the SM \cite{Soares91,WY94}, and in
some extensions of the SM \cite{AI96,AYI97}.
 Much of the theoretical improvements carried out in the
context of the decay $\BbarGAMAXS$ mentioned above can be taken over for
the decay $\BbarGAMAXD$. Like the
former decay, the NLL-improved and
power-corrected decay rate for $\BbarGAMAXD$ rests on firmer theoretical
ground and consequently has
much reduced theoretical uncertainty; in particular, the one
arising from the scale-dependence which in the LL approximation is
notoriously large, becomes drastically reduced in the complete NLL order,
presented here. Hence,
the improved theory for the decay rate would allow to extract more
precisely the CKM parameters from the measured branching ratio 
$\BbarGAMAXD$. Of particular theoretical interest is the ratio of the
branching ratios, defined as
\begin{equation}
\label{dsgamma}
R(d\gamma/s\gamma) \equiv \frac{\langle \BBGAMAXD \rangle}
                           {\langle \BBGAMAXS \rangle}~,
\end{equation}
in which a good part of theoretical uncertainties (such as from $m_t$,
$\mu_b$, ${\cal B}_{sl}$ etc.) cancel. Anticipating this, the ratio
$R(d\gamma/s\gamma)$ was proposed in \cite{ag2} as a good measure of
$\vert V_{td} \vert$. We now calculate this ratio in the NLL 
accuracy, determine its CKM-parametric dependence precisely
and estimate theoretical errors. 

\begin{figure}
\begin{center}
\epsfxsize=10.5cm
\leavevmode
\epsfbox[20 187 573 577]{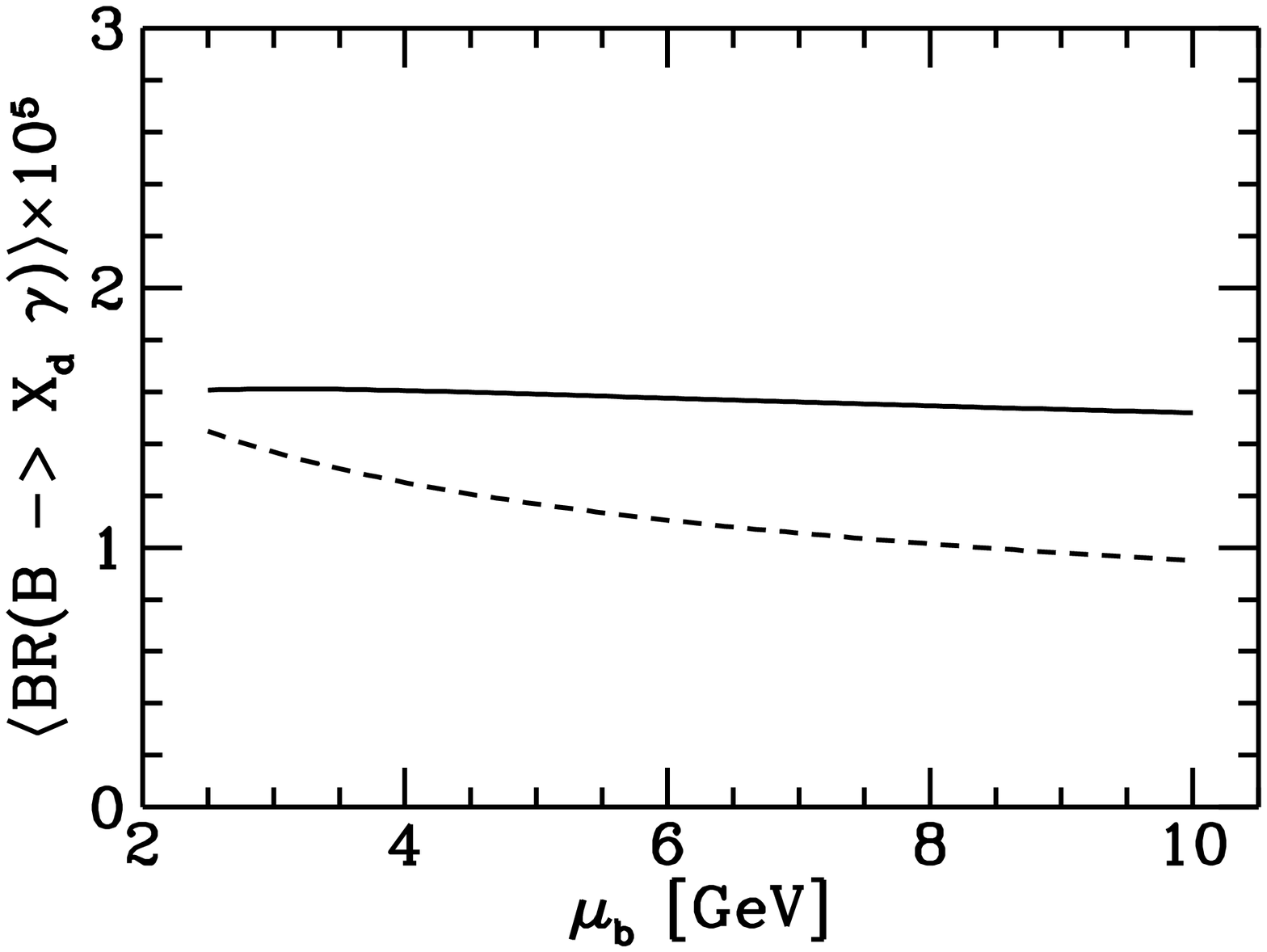}
\end{center}
\caption[]{
Average branching ratio of the processes $\BGAMAXD$ and
$\BbarGAMAXD$, plotted as a function of the scale 
$\mu_b$ for the  central values of the input parameters
$m_b$, $m_c/m_b$, ${\cal B}_{sl}$, $m_t$, $\alpha_{em}$ and $\alpha_s(m_Z)$. 
The solid (dashed) curve shows the NLL (LL) result. \label{fig1}}
\end{figure}

The CP-asymmetry in the decay rates, defined as
\begin{equation}
\label{asymdef}
a_{CP}(B \to X_d + \gamma)\equiv
 \frac{\Gamma(B \to X_d + \gamma)- \Gamma(\overline{B} \to \overline{X_d} +
\gamma)}{\Gamma(B \to X_d + \gamma) + \Gamma(\overline{B} \to
\overline{X_d} +
\gamma)}\end{equation}
has {\it not} so far been calculated in the NLL precision.
We recall that, as opposed to the decay rates $\Gamma(\BbarGAMAXS)$ and
$\Gamma(\BbarGAMAXD)$, which receive contributions starting from
the lowest order, i.e., terms of the form 
$(\alpha^n_s(m_b) \ln^n(m_W/m_b))$, 
the CP-odd  
numerator in eq.~(\ref{asymdef}) is suppressed by an extra
factor $\alpha_s$, i.e., it starts with terms of the form
$\alpha_s(m_b) \,(\alpha_s^n \log^n(m_W/m_b))$. 
To simplify the language in the following, we refer to this statement by 
saying 
that the CP-odd numerator starts at order $\alpha_s$. This
results in a moderate scale dependence of $a_{CP}$,  arising from the
Wilson coefficients which contain a term proportional 
to $\alpha_s \ln (\mu_b/m_b)$ which is not compensated by the matrix
elements in this order.
We show the scale-dependence of $a_{CP}$
numerically by varying the scale $\mu_b$ in the range $2.5 \, \mbox{GeV}
 \leq \mu_b
\leq 10$~GeV. The compensation of this scale dependence
requires the knowledge of the $O(\alpha_s^2)$ contributions
in the matrix elements of the operators in the Wilson product expansion,
which is not yet available.
 However, it is not unreasonable to anticipate that a judicious
choice of the scale $\mu_b$ in $B$ decays may reduce the NLL corrections.
Since the results for the CP-even part, i.e., the denominator in
eq.~(\ref{asymdef}), are known in the LL  approximation, and with the 
help of the present
work now also in the NLL accuracy, this information can be used to 
guess the optimal scale. We make the choice $\mu_b=2.5$~GeV for which
we show that the NLL
corrections in the decay rates become minimal (see Fig.~\ref{fig1}).
 Of course, one can not
insist that this feature must necessarily also hold for $a_{CP}$. 
Not having the benefit of the complete $O(\alpha_s^2)$ calculation for
$a_{CP}$, this particular choice of $\mu_b$ is an educated guess  
based on the inclusive decay rates presented here.

    The branching ratio $\BBbarGAMAXD$ and $a_{CP}$ depend on the parameters
$\rho$ and $\eta$ and this dependence is the principal interest in
measuring these quantities. To estimate them, we vary these parameters
in the range $-0.1 \leq \rho \leq 0.4$ and $0.2 \leq \eta \leq 0.46$, which
are the $95\%$ C.L. ranges allowed by the present fits of the
CKM matrix \cite{aliapctp97}. In addition to this, there are other
well-known parametric dependences inherent to the theoretical framework being
used here,
such as $\alpha_s$, $m_t$, $m_b$, $m_c$, $\alpha_{em}$ 
and ${\cal B}_{sl}$. We estimate the
resulting uncertainty in the branching ratios for
$\BGAMAXD$ and $\BbarGAMAXD$ in an analogous way as has been
done for $\BBbarGAMAXS$ (and its charge conjugate). 
The resulting average
branching ratio $\BRAV=(\BBGAMAXD + \BBbarGAMAXD)/2$ and the
CP rate asymmetry 
$a_{CP}$ are found to be in the range $6.0 \times
10^{-6} \leq \BRAV
 \leq 2.6 \times 10^{-5}$,
and $a_{CP}=(7 - 35)\%$, with most of the dispersion arising due to the
CKM parametric dependence of these quantities. For the central
values of
the fit-parameters \cite{aliapctp97}
$\rho=0.11, \eta=0.33$  
one obtains
$\BRAV = (1.61 \pm 0.12) \times 10^{-5}$ 
and $a_{CP}= (11.5-16.5)\%$, where the errors reflect
the uncertainties stemming from varying 
$\mu_b \in [2.5\,\mbox{GeV},10.0 \, \mbox{GeV}]$ and the
rest of the parameters. The ratio $R(d\gamma/s\gamma)$, defined in
eq.~(\ref{dsgamma}), is largely free of parametric uncertainties;
the residual theoretical error on this quantity is 
small but correlated with the values of $\rho$ and $\eta$.  
Amusingly, the theoretical uncertainty on this ratio almost vanishes  for
the central values of the CKM-fit parameters! (see Fig.~\ref{figrdsgam}). 
 However,
as the presently allowed CKM-domain is large, one can take the 
largest uncertainty $\Delta R(d\gamma/s\gamma)/ R(d\gamma/s\gamma) = \pm
7\%$, which is found for $\eta=0.2$ and $\rho=0.4$, as an 
upper limit on this uncertainty. The ratio $R(d\gamma/s\gamma)$ would then 
provide theoretically the most robust constraint on the CKM parameters.
 For the ratio itself, 
  varying the CKM parameters in the $95\%$ C.L. range, we find
$0.017 \leq R(d\gamma/s\gamma) \leq 0.074$, with the central value being
0.046. Hence, even a first measurement of this ratio will provide a
rather stringent constraint on the $(\rho,\eta)$ domain. We show this
for three assumed values, $R(d\gamma/s\gamma)=0.017,~0.046,~0.074$ taking 
into account the theoretical errors (see Fig.~\ref{ratio}). 

Although the shape of the photon energy spectra in $\BGAMAXS$
and $\BGAMAXD$ is very similar, we think that a measurement of the much 
rarer decay $\BGAMAXD$ should become feasible at future experiments
like CLEO-III and B-factories, because these facilities will allow for 
a good $K/\pi$ discrimination.

 This paper is organized as follows: In section 2,
we discuss the theoretical framework and present the salient features
of the calculation for the decay rates in
$\BbarGAMAXS$ and its charge conjugate process and the CP asymmetry in the
decay rates. In section 3, we work out the corresponding decay rates
and CP asymmetry for $\BbarGAMAXD$, and the ratio $R(d\gamma/s\gamma)$. 
Section 4 contains the
numerical results and we conclude with a summary in section 5.\\

{\bf 2. Decay rates and CP asymmetry in  $\BGAMAXS$ and $\BbarGAMAXS$}\\

  The appropriate framework to incorporate
QCD corrections is that of an effective theory obtained by integrating
out the
heavy degrees of freedom, which in the present context are the top quark
and $W^\pm$ bosons.
 The effective Hamiltonian depends on the underlying theory and
for the SM one has (keeping operators up to dimension 6),
\begin{equation}\label{heffbsg}
{\cal H}_{eff}(b \to s \gamma (+g)) = - \frac{4 G_F}{\sqrt{2}} \lambda_t
        \sum_{i=1}^{8} C_i (\mu) O_i (\mu) ,
\end{equation}
where the operator basis and the 
corresponding Wilson coefficients $C_{i}(\mu)$ can be
seen elsewhere \cite{Misiak96}. The symbol $\lambda_t\equiv V_{tb}V_{ts}^*$
is the relevant CKM factor and $G_F$ is the Fermi coupling constant.

The Wilson coefficients at the renormalization scale
$\mu_b=O(m_b)$ are calculated with the help of
the renormalization group equation whose solution requires the
knowledge of the anomalous dimension matrix in a given order in $\as$
and the matching conditions, i.e., the
Wilson coefficients $C_{i}(\mu=m_W)$, calculated in the complete 
theory to the commensurate order.
 The anomalous dimension matrix in the LL  \cite{Ciuchini}
and the NLL approximation \cite{Misiak96} are known. The NLL
matching conditions
have also been worked out in the meanwhile by several
groups. Of these, the first six
corresponding to the four-quark operators have been derived in
\cite{Buraswc}, and the remaining two, $C_7(\mu=m_W)$ and
$C_8(\mu=m_W)$, were worked out in \cite{Yao94} and confirmed in
\cite{Greub97}, \cite{BKP97} and \cite{CDGG97}.. In addition,
the NLL corrections to the matrix elements have also been calculated.
Of these, the Bremsstrahlung corrections were obtained in
\cite{ag1,ag2} in the truncated basis (involving the operators
$O_1$, $O_2$, and $O_7$) and 
subsequently in the complete operator basis  \cite{ag95,Pott95}.
The NLL virtual corrections were completed in
\cite{GHW96}. This latter contribution  plays a key role in reducing
the scale-dependence of the LL inclusive decay
width. All of these pieces have been combined to get the NLL decay width
$\GbarGAMAXS$ and the details are given in the literature
\cite{Misiak96}-\cite{BKP97}.

We
recall that the operator basis in ${\cal H}_{eff}$ is in fact larger than
what is shown in eq. (\ref{heffbsg}) in which operators multiplying 
the small CKM factor
$\lambda_u\equiv V_{ub}V_{us}^*$ have been neglected.
If the interest is in calculating the CP asymmetry, then
they have to be put back. Doing this, and
using the unitarity relation $\lambda_c=-\lambda_t-\lambda_u$, the
effective Hamiltonian reads
\begin{eqnarray}
\label{heffbsgcp}
{\cal H}_{eff}(b \to s \gamma (+g))  &=& 
-  \frac{4G_F}{\sqrt{2}}
\left\{ \lambda_{t} \left[ C_7(\mu)O_{7}(\mu)+
C_8(\mu)O_{8}(\mu)+C_1(\mu)O_1(\mu)+C_2(\mu)O_{2}(\mu)\right] \right.
\nonumber \\ 
&& \left. \hspace{0.7cm} 
-\lambda_u \left[C_1(\mu)(O_{1u}(\mu)-O_1(\mu))+
C_{2}(\mu)(O_{2u}(\mu)-O_{2}(\mu))\right]+ \cdots \right\} \,.
\nonumber \\
\end{eqnarray}
In this equation terms 
proportional to the small Wilson coefficients $C_3,...,C_6$
are dropped as
indicated by the 
ellipses. The relevant operators are defined as:
\begin{eqnarray}
\label{operators}
O_{1}(\mu)&=&(\bar{s}_{L}\gamma_{\mu}T^a c_{L})
(\bar{c}_{L}\gamma^{\mu}T^a b_{L}) \quad , \quad
O_{1u}(\mu)=(\bar{s}_{L}\gamma_{\mu}T^a u_{L})
(\bar{u}_{L}\gamma^{\mu}T^a b_{L}) ~, \nonumber \\
O_{2}(\mu)&=&(\bar{s}_{L}\gamma_{\mu} c_{L})
(\bar{c}_{L}\gamma^{\mu} b_{L})\phantom{T^aT^a} \quad , \quad
O_{2u}(\mu)=(\bar{s}_{L}\gamma_{\mu} u_{L})
(\bar{u}_{L}\gamma^{\mu} b_{L}) ~, \nonumber \\
O_{7}(\mu)&=&\frac{e}{16\pi^2}m_{b}(\mu)(\bar{s}_{L}\sigma_{\mu
\nu}b_{R}) F^{\mu \nu} \quad , \quad
O_8(\mu)=\frac{g}{16\pi^2}m_b(\mu)(\bar{s}_{L}T^{a}
\sigma_{\mu \nu}b_{R})G^{a \mu \nu}~.
\end{eqnarray}
Note that the Wilson coefficients in eq. (\ref{heffbsgcp}) are exactly
the same as those in eq. (\ref{heffbsg}). Moreover, the matrix 
elements $<s\gamma|O_{iu}|b>$ and $<s\gamma g|O_{iu}|b>$  
of the
additional operators $O_{1u}$ and $O_{2u}$ are obtained from
those of $O_1$ and $O_2$ by obvious replacements. 

For our intent and purpose, we write the amplitudes for the processes
$b \to s
\gamma$ and $b \to s \gamma g$ in a form where the dependence
on the CKM matrix elements is manifest. The amplitude for the first
process (including the virtual corrections) can be written as
 \begin{eqnarray}
A(b\to s\gamma)&=&-\frac{4G_F}{\sqrt{2}}<s\gamma|O_7|b>_{tree}
D(b\to s \gamma)~, \nonumber \\
D(b\to s \gamma)&=&\lambda_t(A_R^t+iA_I^t)+\lambda_u(A_R^u+iA_I^u)~.
\end{eqnarray}
It is straightforward to construct
the real functions $A_R^t, A_I^t,A_R^u$ and  $A_I^u$
from the expressions for the virtual correction in ref. \cite{GHW96}
and the NLL Wilson coefficients in ref. \cite{Misiak96}.
The amplitude of the charge conjugate decay $\overline{b}
 \to \overline{s}\gamma$
decay is then:
\begin{eqnarray}
A(\overline{b} \to \overline{s}
 \gamma)&=&-\frac{4G_F}{\sqrt{2}}<s\gamma|O_7|b>_{tree}
D(\bar{b}\to \bar{s} \gamma)~,                                     
\nonumber \\
D(\bar{b}\to \bar{s}
\gamma)&=&\lambda_t^*(A_R^t+iA_I^t)+\lambda_u^*(A_R^u+iA_I^u) 
\,.
\end{eqnarray}
The  decay rate for the process $b \to s \gamma$ then reads
\footnote{Note that we have absorbed the factor 
$F=1- (8\alpha_s)/(3\pi)$, present in ref. \cite{GHW96}, 
into the term $A_R^t$.}
\begin{eqnarray}
\label{dsquare}
\Gamma(b \to s \gamma) &=&
\frac{m_{b}^5 \, G_F^2 \, \alpha_{em}}{32\pi^4} \, |D(b\to s \gamma)|^2~,
\nonumber \\
|D(b\to s \gamma)|^2 &=& |\lambda_t|^2 \left[ (A_R^t)^2+
(A_I^t)^2\right]+|\lambda_u|^2 \left[ (A_R^u)^2+(A_I^u)^2 \right] + 
\nonumber \\
&& 2Re(\lambda_t^*\lambda_u) \left[ A_R^tA_R^u+A_I^tA_I^u \right] -
2Im(\lambda_t^*\lambda_u) \left[ A_R^tA_I^u-A_I^tA_R^u \right] \,.
\end{eqnarray}
The order $\alpha_s^2$ terms, which are generated when inserting 
the explicit expressions for the functions
$A_R^t$, $A_I^t$, $A_R^u$, and $A_I^u$, are understood to be
discarded.
The corresponding expression for the $\overline{b} \to 
\overline{s} \gamma$ decay can be
 obtained from the preceding equation
 by changing the sign of the term proportional to
$Im(\lambda_t^*\lambda_u)$.

An analogous expression for the decay width 
 $\Gamma(b\to s \gamma g)$
of the Bremsstrahlung
process,
where the CKM
dependence is explicit, is also easily obtained from the literature
\cite{ag1,ag2,ag95,Pott95,GHW96}. 
To get rid of the infrared singularity for $E_\gamma \to 0$, we included
the virtual photonic correction to the process $b \to s g$, as discussed
in these references. Another possibility, which was suggested in ref. 
\cite{Misiak96}, is to define the branching ratio in such a way that the
 photon energy $E_\gamma$ has to be larger than some minimal value
$E_\gamma^{min}$.

 It is customary to express the
 branching ratio  $\BBbarGAMAXS$ in terms of the measured
semileptonic  branching ratio ${\cal B} (B \to X\ell \nu_\ell)$,
\begin{equation}
\label{brbsgsm}
\BBbarGAMAXS = \frac{\GbarGAMAXS}{\Gamma_{sl}}
\, {\cal B} (B \to X\ell \nu_\ell).
\end{equation}
The expression for $\Gamma_{sl}$ (including radiative corrections)
can be seen  in refs.~\cite{Cabibbo}.

In addition to the perturbative QCD improvements discussed above, also the
leading power corrections, which start in $1/\mb^2$, have been
calculated to the
decay widths appearing in the numerator and denominator of
eq.~(\ref{brbsgsm}) \cite{georgi,manoharwise,FLS94}.
The power corrections in the numerator have been obtained assuming that the
decay $\BbarGAMAXS$ is dominated by the magnetic moment operator $O_7$.
Writing this correction in an obvious notation as
\begin{equation}
\frac{\Gamma(\BbarGAMAXS)}{\Gamma^{0}(\BbarGAMAXS)} = 1 + 
\frac{\delta_b}{m_b^2},
\end{equation}
one obtains $\delta_b =1/2 \lambda_1 -9/2 \lambda_2$, where $\lambda_1$ and
$\lambda_2$ are, respectively, the kinetic energy and magnetic moment
parameters of the theoretical framework based on heavy quark expansion.
 Using $\lambda_1 =-0.5
~\mbox{GeV}^2$ and $\lambda_2
=0.12 ~\mbox{GeV}^2$, one gets $\delta_b/\mb^2 \simeq -4\%$. However, 
the leading order $(1/m_b^2)$ power
corrections in the heavy quark expansion proportional to $\lambda_1$ are 
identical
in the inclusive decay rates  $\Gamma(\BbarGAMAXS)$ and $\Gamma(B \to X \ell
\nu_\ell)$. The corrections proportional to
$\lambda_2$ differ only marginally. Thus, including or
neglecting the $1/m_b^2$ corrections makes a difference of only $1\%$ in
$\BBbarGAMAXS$.

The power corrections proportional to $1/m_c^2$, resulting from the
interference of the operator $O_2$ (and $O_1$) with
 $O_7$ in $\BbarGAMAXS$, have also
been
worked out \cite{Voloshinbsg97,powermc,BIR97}.
Expressing this symbolically as
\begin{equation}
\frac{\Gamma(\BbarGAMAXS)}{\Gamma^{0}(\BbarGAMAXS)} = 1 + 
\frac{\delta_c}{m_c^2}~,
\end{equation}
one finds $\delta_c/m_c^2 \simeq + 0.03$ \cite{BIR97}.

It is convenient to express the branching ratio 
for $\BbarGAMAXS$ in a form where the dependence on the CKM
matrix factors is manifest:
\begin{eqnarray}
\label{brxs}
\BBbarGAMAXS=\frac{|\lambda_t|^2}{|V_{cb}|^2}D_{t}+
\frac{|\lambda_u|^2}{|V_{cb}|^2}D_{u}+
\frac{Re(\lambda^{*}_t\lambda_u)}{|V_{cb}|^2}D_{r}+
\frac{Im(\lambda^{*}_t\lambda_u)}{|V_{cb}|^2}D_{i} \, .
\end{eqnarray}
The quantities $D_a$ $(a=t,u,r,i$), which
depend on various input parameters such  as
$m_t,m_b,m_c,\mu_b$ and $\as$, 
are calculated numerically and listed in Table 1.
The averaged branching ratio $\langle \BBGAMAXS \rangle$
is obtained  by discarding the last term on the right hand side of 
eq. (\ref{brxs}).
The CP-violating rate asymmetry has been defined earlier. In
terms of the functions $D_a$ it can be expressed as:
\begin{equation}
\label{acpnlo}
a_{CP}(\BGAMAXS)=-\frac{Im(\lambda_t^*\lambda_u)D_i}
{|\lambda_t|^2D_{t}+|\lambda_u|^2D_{u}+
Re(\lambda^{*}_t\lambda_u)D_{r}}~.
\end{equation}
Since the function $D_i$ in the numerator in eq. (\ref{acpnlo})
only starts at order $\alpha_s$,  
the complete NLL expression for $a_{CP}$ requires
$D_i$ up to and including the $O(\alpha_s^2)$ term which is not known.
Hence, in the LL approximation, a consistent definition
of $a_{CP}$ is the one in which only the LL result for the denominator
is retained, i.e., in this approximation one
should drop terms proportional to $D_u$ and $D_r$ and keep only the
LL result for $D_t$, which is denoted as
$D_t^{(0)}$ in the following. 
The expression for $a_{CP}$
in this approximation  then reduces to
\begin{equation}
\label{acpll}
a_{CP}(\BGAMAXS)=-\frac{Im(\lambda_t^*\lambda_u)D_i}
{|\lambda_t|^2 \, D_t^{(0)}} \,.
\end{equation}
This is what we shall use in the numerical estimates of $a_{CP}$.
Note that $D_t^{(0)}$, which is also shown in Table 1, is proportional
to the square of the LL Wilson coefficient $C_7^{0,\,{\rm eff}}(\mu_b)$.
To be precise, this Wilson coefficient is obtained from the Wilson
coefficients $C_i$ ($i=1,...,8$) at the matching scale $\mu=m_W$ 
by using the 1-loop
expression for $\alpha_s(\mu)$ in the renormalization group
evolution. Moreover, notice that
the power
corrections, which are contained
 in the functions $D_t$ and $D_r$ in the NLL
branching ratio (\ref{brxs}), 
drop out in the LL expression for $a_{CP}$.

In the Wolfenstein parametrization \cite{Wolfenstein},
which will be used in the numerical analysis,
the CKM matrix is determined in terms of the four parameters
$A, \lambda=\sin \theta_C$, $\rho$ and $\eta$, and one can express the 
quantities $\l_t$, $\l_u$ and $|V_{cb}|^2$ in the above equations as
\cite{BLO94}
(neglecting terms of $O(\l^6))$:
\begin{equation}
\label{wolfenlambda}
\l_u = A \lambda^4 \, (\rho - i \eta),
~~~\l_t =  -A \l^2 \left( 1-\frac{\l^2}{2}+\l^2(\rho-i\eta)\right),
~~~\l_c=-\l_u - \l_t,
~~~|V_{cb}|^2=A^2 \l^4 \,.
\end{equation}

{\bf 3. Decay rates and CP asymmetry in $\BGAMAXD$ and $\BbarGAMAXD$}\\

 In complete analogy
with the $\BbarGAMAXS$ case discussed earlier,
the relevant set of dimension-6 operators for
the processes $b \to d \gamma$ and $b \to d \gamma g$
can be written as
\begin{eqnarray}
\label{heffbdgcp}
{\cal H}_{eff}(b \to d \gamma (+g))  &=& 
-  \frac{4G_F}{\sqrt{2}}
\left\{ \xi_{t} \left[ C_7(\mu)O_{7}(\mu)+
C_8(\mu)O_{8}(\mu)+C_1(\mu)O_1(\mu)+C_2(\mu)O_{2}(\mu)\right] \right.
\nonumber \\ 
&& \left. \hspace{0.7cm} 
-\xi_u \left[C_1(\mu)(O_{1u}(\mu)-O_1(\mu))+
C_{2}(\mu)(O_{2u}(\mu)-O_{2}(\mu))\right]+ \cdots \right\} \, ,
\nonumber \\
\end{eqnarray}
where $\xi_{j} = V_{jb} \, V_{jd}^{*}$ with $j=u,c,t$. The operators
are the same as in eq. (\ref{operators}) up to the obvious
replacement of the $s$-quark field by the $d$-quark field.  
Moreover, the matching conditions $C_i(m_W)$
and the solutions
of the RG equations, yielding $C_i(\mu_b)$, coincide
with those needed for the process $b \to s \gamma (+g)$.
The power
corrections in $1/m_b^2$ and $1/m_c^2$ (besides the CKM factors)
are also the same for $\Gamma(\BbarGAMAXD)$
and $\Gamma(\BbarGAMAXS)$. However, the so-called long-distance contributions
from the intermediate $u$-quark in the penguin loops are {\it different}
in the
decays $\BbarGAMAXS$ and $\BbarGAMAXD$.
These are  suppressed in the decays $\BbarGAMAXS$ due to the unfavorable CKM
matrix elements. In $\BbarGAMAXD$, however, there is no CKM-suppression 
and one
has to include the long-distance intermediate $u$-quark contributions. It
must be stressed that
there is no spurious enhancement 
of the form $\ln (m_u/\mu_b)$ 
in the perturbative contribution to the matrix elements
$<\overline{X_d}\gamma|O_{iu}|\overline{B}>$ ($i=1,2$)
as shown by the explicit calculation in \cite{GHW96} and also
discussed more recently in \cite{ARS97}. In other words, the limit $m_u
\to 0$ can be taken safely. 
The non-perturbative contribution generated by the $u$-quark loop
can
only be modeled at present. In this context, we recall that estimates
based on the vector meson dominance indicate that these
contributions are small \cite{deshpande95}. Estimates of the long-distance
contributions in exclusive decays $B \to \rho \gamma$ and $B \to \omega
\gamma$ in the Light-Cone QCD sum rule approach put the corresponding
corrections somewhere around $O(15\%)$ for the
charged ($B^\pm$) decays and much smaller $O(5\%)$ for the neutral $B$
decays \cite{ab95,ksw95}. Model estimates based on final state interactions
likewise give small long-distance contribution for the exclusive radiative
$B$ decays \cite{DGP96}. To take this uncertainty into account,
we add an error   proportional to $D_t$ in LL approximation, viz. 
$\pm 0.1D_t^{(0)}$, in the numerical estimate of the function $D_r$
when calculating the branching ratio $\BBbarGAMAXD$ \cite{deshpande95}.

In analogy to eq. (\ref{brxs}) 
the branching ratio $\BBbarGAMAXD$ 
in the SM  can be written as
\begin{eqnarray}
\label{brxd}
\BBbarGAMAXD=\frac{|\xi_t|^2}{|V_{cb}|^2}D_{t}+
\frac{|\xi_u|^2}{|V_{cb}|^2}D_{u}+
\frac{Re(\xi^{*}_t\xi_u)}{|V_{cb}|^2}D_{r}+
\frac{Im(\xi^{*}_t\xi_u)}{|V_{cb}|^2}D_{i} \, ,
\end{eqnarray}
where the functions $D_a$ $(a=t,u,r,i)$ 
are the same as in eq. (\ref{brxs}).
While these functions (or some combinations thereof)
were obtained in partial NLL approximation some time ago
\cite{ag2}, the complete NLL results
are presented here for the first time.
For numerical values
of these functions we refer to Table 1.
An expression for the averaged branching ratio $\BRAV$ is
obtained by dropping the last term on the right hand side in eq.
(\ref{brxd}). 

As the branching ratio $\langle \BBGAMAXS \rangle$ is 
very well approximated
by $\langle \BBGAMAXS \rangle= |\l_t|^2 D_t/|V_{cb}|^2$,
the ratio $R(d\gamma/s\gamma)$, defined in eq.~(\ref{dsgamma}), can
be expressed as follows:
\begin{equation}
\label{bsgamckm}
R(d\gamma/s\gamma) = \frac{|\xi_t|^2}{|\l_t|^2}
+ \frac{D_u}{D_t} \, \frac{|\xi_u|^2}{|\l_t|^2}
+ \frac{D_r}{D_t} \, \frac{Re(\xi_t^* \xi_u)}{|\l_t|^2} \, .
\end{equation}
The leading term $|\xi_t|^2/|\l_t|^2$ 
in eq.~(\ref{bsgamckm}) is obviously independent of any 
dynamical uncertainties;
the subleading terms proportional to $D_u/D_t$ and $D_r/D_t$ are still
uncertain by almost a factor 2, but numerically small compared to unity
(see Table 1).
Also, as we shall see in the next section, the allowed
values of the CKM parameters provide a further suppression of these terms. 
Hence, the overall uncertainty in $R(d\gamma/s\gamma)$ is small.

Using again the LL expression for the denominator in eq.
(\ref{asymdef}), the CP rate asymmetry can be written as
\begin{equation}
\label{acpdll}
a_{CP}(\BGAMAXD)=-\frac{Im(\xi_t^*\xi_u)D_i}
{|\xi_t|^2 \, D_t^{(0)}} \, ,
\end{equation}
where $D_t^{(0)}$ stands for the LL expression of $D_t$. 
In the numerical analysis, we will use the following expressions 
for 
the quantities $\xi_j$ in eqs. (\ref{brxd}) and (\ref{acpdll})
(neglecting terms of $O(\l^7)$):
\begin{equation}
\label{wolfenxi}
\xi_u = A \, \lambda^3 \, (\bar\rho - i \bar\eta),
~~~\xi_t =  A \, \lambda^3 \, (1 - \bar\rho + i \bar\eta) ,
~~~\xi_c=-\xi_u - \xi_t,
\end{equation}
with $\bar\rho = \rho (1-\l^2/2)$ and
$\bar\eta = \eta (1-\l^2/2)$ \cite{BLO94}.
Note that all three CKM-angle-dependent quantities
$\xi_j$ start at order $\lambda^3$.
Inserting these expressions into eq. (\ref{acpdll}),
a simple form for the CP rate asymmetry is obtained: 
\begin{equation}
a_{CP}(\BGAMAXD)= 
\frac{D_i \bar\eta}{D_t^{(0)} 
\left[(1-\bar\rho)^2 + \bar\eta^2\right] }\, .
\end{equation}

{\bf 4. Numerical Estimates of branching ratios and CP asymmetries}\\

We now proceed to the numerical analysis of our results. Based on present
measurements and theoretical estimates, we take the following values for
the input parameters:
$\alpha_s(M_Z)=0.118\pm 0.003$, $m_b=4.8\pm 0.15$ GeV,
$m_c/m_b=0.29\pm 0.02$, $m_t \equiv
 m_t(\mbox{pole})=(175\pm 6)$~GeV (corresponding to
$\overline{m_t(m_t)}=(168\pm 6)$~GeV), ${\cal B}_{sl}=(10.49 \pm 0.46)\%$,
$\alpha_{em}^{-1}=(130.3 \pm 2.3)$. 
For the CKM matrix elements we note that
the parameters A and $\lambda$ are rather well determined. The
parameters $\rho$ and $\eta$ are constrained from unitarity fits. The
updated fits, taking into account also the lower bound on the mixing-induced
mass difference ratio $\Delta M_s/\Delta M_d > 20.4$ yield (at $\pm 1 
\sigma$) \cite{aliapctp97},
\begin{eqnarray}
\nonumber
&&A=0.81 \pm 0.057,  \hspace{2cm} \lambda=0.22 \\
&& \eta=0.33 \pm 0.065,
\hspace{1cm}
 \rho =0.11^{+0.14}_{-0.11},
\end{eqnarray}
where $\lambda$, being very accurately measured, was fixed to the
value shown. Note that the allowed range of $\rho$ is now asymmetric with
respect to $\rho=0$ due to the mentioned bound on $\Delta M_s/\Delta M_d$,
which removes large negative-$\rho$ values. We also note that 
the recent CKM fits reported in \cite{Parodietal} yield an identical
range for $\eta$ but they find $\rho =0. 156 \pm 0.090$, which is more 
restrictive for the lower bound on $\rho$ than the analysis in
\cite{aliapctp97}, that we use here.

%
%
\begin{table}[htb]
\begin{center}
\begin{tabular}{|r|r|r|r|r|}  \hline
& $\mu_b=2.5\, \mbox{GeV}$ & $\mu_b=5 \, \mbox{GeV}$ & 
$\mu_b=10 \, \mbox{GeV}$ & $m_c/m_b$ \\ \hline
$D_t^{(0)}/\l^4$ &  0.131 &  0.106 &  0.086 & 0.27 \\ \hline
$D_t^{(0)}/\l^4$ &  0.142 &  0.114 &  0.093 & 0.29 \\ \hline
$D_t^{(0)}/\l^4$ &  0.155 &  0.125 &  0.101 & 0.31 \\ \hline
$D_t/\l^4$ &        0.150 &  0.147 &  0.140 & 0.27 \\ \hline
$D_t/\l^4$ &        0.155 &  0.154 &  0.147 & 0.29 \\ \hline
$D_t/\l^4$ &        0.161 &  0.163 &  0.157 & 0.31 \\ \hline
$D_u/\l^4$ &        0.015  & 0.011 &  0.009 & 0..27 \\ \hline
$D_u/\l^4$ &        0.016  & 0.012 &  0.009 & 0.29 \\ \hline
$D_u/\l^4$ &        0.016  & 0.012  & 0.009 & 0.31 \\ \hline
$D_r/\l^4$ &       $-0.033$ & $-0.021$ &  $-0.014$ & 0.27 \\ \hline
$D_r/\l^4$ &       $-0.043$ & $-0.028$ &  $-0.019$ & 0.29 \\ \hline
$D_r/\l^4$ &       $-0.055$ & $-0.036$ &  $-0.025$ & 0.31 \\ \hline
$D_i/\l^4$ &        0.056 &  0.039 &  0.028 & 0.27 \\ \hline
$D_i/\l^4$ &        0.062 &  0.042 &  0.031 & 0.29 \\ \hline
$D_i/\l^4$ &        0.068 &  0.047 &  0.034 & 0.31  \\ \hline
\end{tabular}
\end{center}
\caption{Values of the NLL functions 
$D_t, ~D_u,~D_r,~D_i$ (divided by $\l^4$) for the
indicated values of the scale parameter $\mu_b$ and the quark mass ratio
$m_c/m_b$. Also tabulated are the values for the LL function 
$D_t^{(0)}/\lambda^4$. } \label{table1}
\end{table}

In Table 1, we give  the values of the functions
$D_t$, $D_u$, $D_r$ and $D_i$, evaluated  for the central values of the
parameters $\alpha_s(m_Z)$, $m_t$, $m_b$, $\alpha_{em}$ 
and the semileptonic branching ratio
${\cal B}_{sl}$. 
The other two parameters $m_c/m_b$ and $\mu_b$ are varied as indicated.
We note that
the renormalization scale dependence of $D_t$ is significantly
reduced in the NLL compared to the LL result $D_t^{(0)}$. 
As $D_u$, $D_r$, and $D_i$ start at order $\alpha_s$ only, their
$\mu_b$
dependence
is more significant, but their contribution to
the branching ratio is rather small.\\

For the values of the input parameters given above,
the theoretical branching ratio for the decay  $\BbarGAMAXS$
in the SM is calculated by us as 
$\BBbarGAMAXS = (3.50 \pm 0.32)\times 10^{-4}$. This is to be compared
with the recent result in ref. \cite{BG98}, where the central value
$3.46 \times 10^{-4}$ is quoted for the case in which the factor
$1/\Gamma_{sl}$ is -- like in the present work -- expanded in $\alpha_s$.
Taking into account, that we use $|\l_t/V_{cb}|^2=0.96$
in the present CKM framework, whereas in ref. \cite{BG98} 
a value $0.95$ was used for the
same quantity, the results here and in \cite{BG98} are in agreement.
To calculate $a_{CP}$ in the decay rates for  $B \to X_s + \gamma $
and its charge conjugate $\BbarGAMAXS$, we shall use the
LL approximation (\ref{acpll}) for $a_{CP}$.
For the central
values of the parameters we obtain (neglecting corrections of $O(\l^2)$):
$a_{CP}(B \to X_s + \gamma)=-(2.11)\eta
\%$, which gives for $\eta = 0.33 \pm 0.13$ the
 following prediction for the decay rate asymmetry:
$a_{CP}(B \to X_s + \gamma)=-(0.70\pm 0.28)\%$. 
Note that these numbers correspond to the preferred scale $\mu_b=2.5$
GeV. For $\mu=5$ GeV, the asymmetry would be 
$a_{CP}(B \to X_s + \gamma)=-(0.59\pm 0.25)\%$. 
Thus, the direct CP asymmetry in  
$\BbarGAMAXS$ 
in the standard model turns out to be too small
to be measurable.\\

We now discuss the decay $\BbarGAMAXD$ and the CP conjugated
process $\BGAMAXD$. 
The averaged branching ratio $\BRAV$ strongly depends on the CKM parameters
$\rho$, $\eta$.
Taking the central values of the parameters $\alpha_s(M_Z)$, $m_b$,
$m_c/m_b$, $m_t$, ${\cal B}_{sl}$, $\alpha_{em}$ 
and $\mu_b=2.5$ GeV, one obtains the following prediction:
\begin{eqnarray}
&&\BRAV = 2.43 \, \left[(1-\bar\rho)^2+
\bar\eta^2-0.35(1-\bar\rho)+0.07\right] \times 10^{-5}~, \nonumber\\
&& \simeq 1.61 \times 10^{-5} ~~~[\mbox{for}~(\rho ,\eta)=(0.11,0.33),~
\mbox{or}~(\bar{\rho},\bar{\eta})=(0.107,0.322)]\, .
\end{eqnarray}
 In comparison,
the result in the LL approximation for $\BRAV$,
for the same values of the parameters  is:
$\BRAV=1.61 \, \left[(1-\bar\rho)^2+\bar\eta^2\right] \times 10^{-5}$. 
This gives,
$\BRAV= 1.45 \times 10^{-5}$ for $(\rho ,\eta)=(0.11,0.33)$.
The difference between the LL and NLL results is 
$\sim 10\%$, increasing the branching ratio in the NLL case
(see Fig.~\ref{fig1} for $\mu_b=2.5$ GeV). The scale $(\mu_b)$-dependence
of $\BRAV$ in the LL and NLL accuracy is shown in Fig.~\ref{fig1}, fixing
all other parameters to their central values. 

\begin{figure}
\begin{center}
\epsfxsize=10.5cm
\leavevmode
\epsfbox[20 187 573 577]{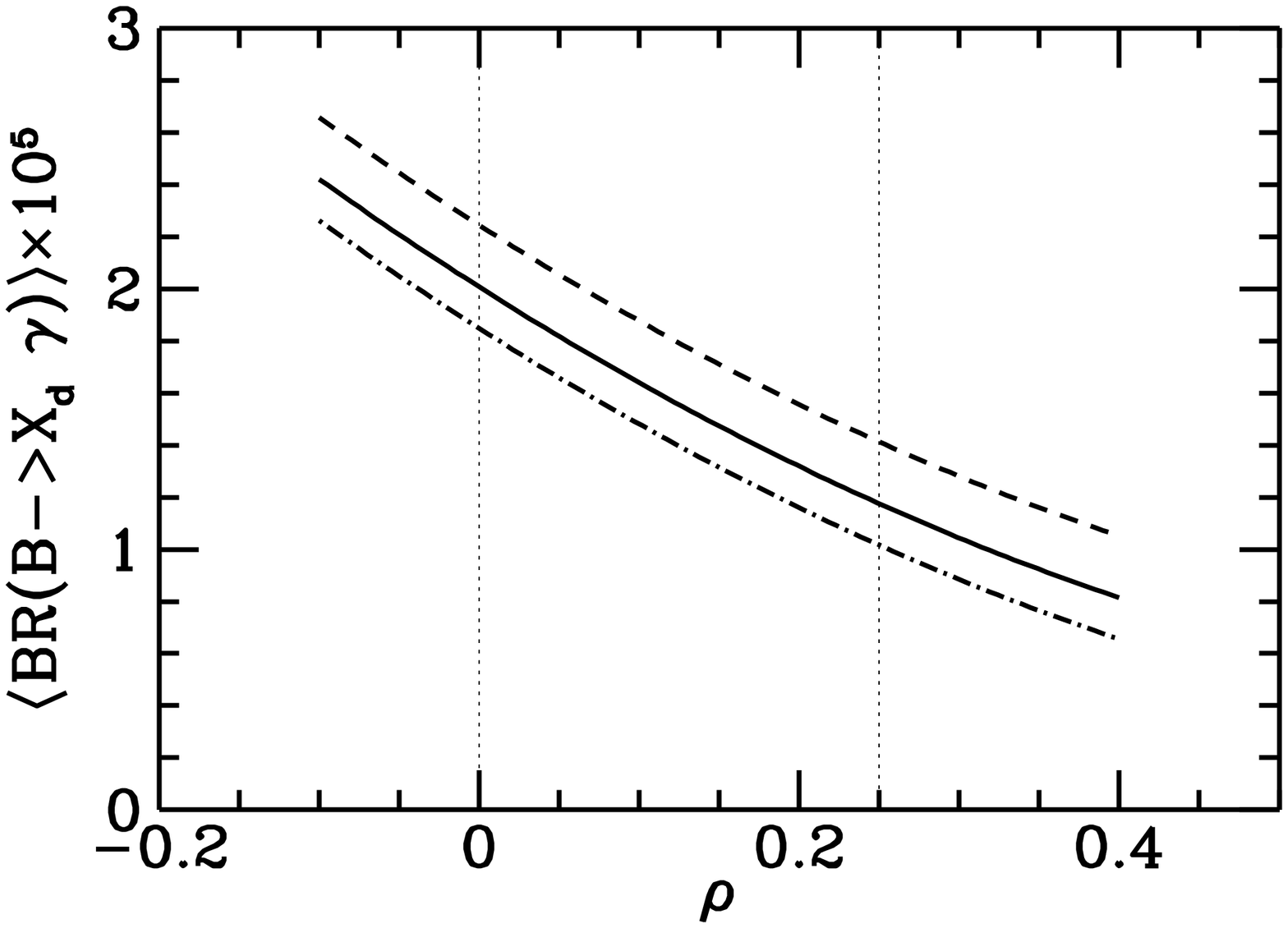}
\end{center}
\caption[]{
The $\rho$ dependence of the average branching ratio for
$\BGAMAXD$ and $\BbarGAMAXD$ is shown 
for different values of $\eta$:
$\eta=0.46$ (dashed curve); $\eta=0.33$(solid curve); $\eta=0.20$ 
(dash-dotted curve).
All three curves correspond to $\mu_b=2.5$ GeV and to the  central values 
of the input parameters
$m_b$, $m_c/m_b$, ${\cal B}_{sl}$, $m_t$, $\alpha_{em}$ and $\alpha_s(m_Z)$.
The vertical lines show the $\pm 1 \sigma$ range for $\rho$ from the
CKM fits \protect\cite{aliapctp97}.
\label{fig2}}
\end{figure}

In Fig.~\ref{fig2} we give the $\rho$ dependence of the branching ratio
$\BRAV$ for $\eta =0.20, 0.33$ and  0.46, using 
$\mu_b=2.5$ GeV and
the central values for all
other parameters. We note that the dependence on $\eta$ is not very marked.
The branching ratio is largest for the smallest allowed value of $\rho$
(taken here as $\rho =-0.10$)
and the largest allowed value of $\eta$ (assumed here as $\eta=0.46$), and
may reach a value of $2.6 \times 10^{-5}$.
The minimum value of $\BRAV$ in the SM is estimated as $6.0 \times
10^{-6}$.

 The ratio $R(d\gamma/s\gamma)$ in eq. (\ref{bsgamckm})
can be expressed in terms of the
CKM parameters $\bar{\rho}$ and $\bar{\eta}$ as follows 
(expanding $1/|\l_t|^2$ in powers of $\l$):
\begin{eqnarray}
\label{ratiosd}
R(d\gamma/s\gamma) &=& \lambda^2 [1+\lambda^2(1-2 \bar{\rho})] \,
\left[(1-\bar{\rho})^2 + \bar{\eta}^2 + \frac{D_u}{D_t} (\bar{\rho}^2 + 
\bar{\eta}^2) + \frac{D_r}{D_t} (\bar{\rho}(1 - \bar{\rho})
 - \bar{\eta}^2)\right] \, ,
\nonumber\\
&\simeq & 0.046 ~~~[\mbox{for}~(\rho ,\eta)=(0.11,0.33),~  
\mbox{or}~(\bar{\rho},\bar{\eta})=(0.107,0.322)]\, .
\end{eqnarray} 
   
 Our prediction for the direct CP asymmetry
$a_{CP}(\BGAMAXD)$, based on the LL result (\ref{acpdll})
 and for the  central
values of the input parameters, is:
\begin{eqnarray}
&&a_{CP}(\BGAMAXD)(\mu_b =2.5\, 
\mbox{GeV})=\frac{0.44\bar\eta}{(1-\bar\rho)^2+\bar\eta^2}~, \nonumber\\
&& \simeq  0.16 
~~~[\mbox{for}~(\rho ,\eta)=(0.11,0.33),~
\mbox{or}~(\bar{\rho},\bar{\eta})=(0.107,0.322)]\, .
 \end{eqnarray}
The scale dependence of this result is
as follows: $a_{CP}(\mu_b= 5 \, \mbox{GeV}) 
\simeq 0.13$ and $a_{CP}(\mu_b=10 \, \mbox{GeV}) \simeq 0.12$.
As argued earlier,
we prefer $\mu_b=2.5$~GeV to estimate $a_{CP}$, as for this choice of the scale
the NLL corrections in the decay rates are small.

\begin{figure}
\begin{center}
\epsfxsize=10.5cm
\leavevmode
\epsfbox[20 187 573 577]{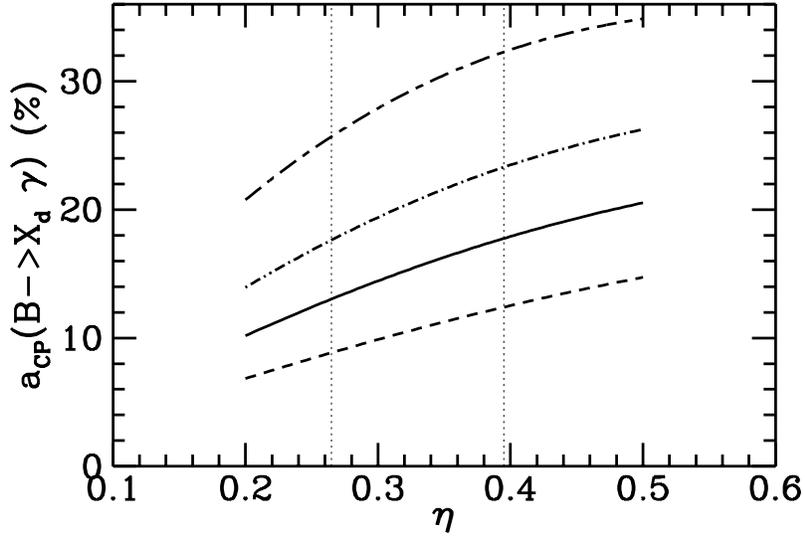}
\end{center}
\caption[]{
$\eta$ dependence of the CP rate asymmetry $a_{CP}(\BGAMAXD)$
for different values of $\rho$:
$\rho=-0.1$ (dashed curve); $\rho=0.11$ (solid curve); 
$\rho=0.25$ (dash-dotted curve),
$\rho=0.4$ (long-short dashed curve).
All four curves correspond to $\mu_b=2.5$ GeV and to the central values 
of the input parameters
$m_b$, $m_c/m_b$, ${\cal B}_{sl}$, $m_t$, $\alpha_{em}$ 
and $\alpha_s(m_Z)$.
The vertical lines show the $\pm 1 \sigma$ range for $\eta$ from the
CKM fits \protect\cite{aliapctp97}.
\label{fig3}}
\end{figure}

In Fig.~\ref{fig3} we show the $\eta$ dependence of the direct CP rate 
asymmetry for
the $\BbarGAMAXD$ decay for $\rho = -0.10, 0.11, 0.25$ and $0.40$, using
again $\mu_b=2.5$ GeV and the
central values of all other parameters. The smallest value of
$a_{CP}(\BGAMAXD)$ is $7\%$ (for $\rho=-0.10$ and $\eta=0.20$) and may
reach as high a value as $35\%$ (for  $\rho=0.4$ and $\eta=0.46$),
as can be seen in Fig.~\ref{fig3}.
We want to stress that the $\rho$-dependence of the
branching ratio $\BRAV$ and both the  $\rho$- and
$\eta$-dependence of $a_{CP}(\BGAMAXD)$ are very marked. Hence,
their measurements will help to determine these parameters more precisely.\\

\begin{figure}
\begin{center}
\epsfxsize=10.5cm
\leavevmode
\epsfbox[20 187 573 577]{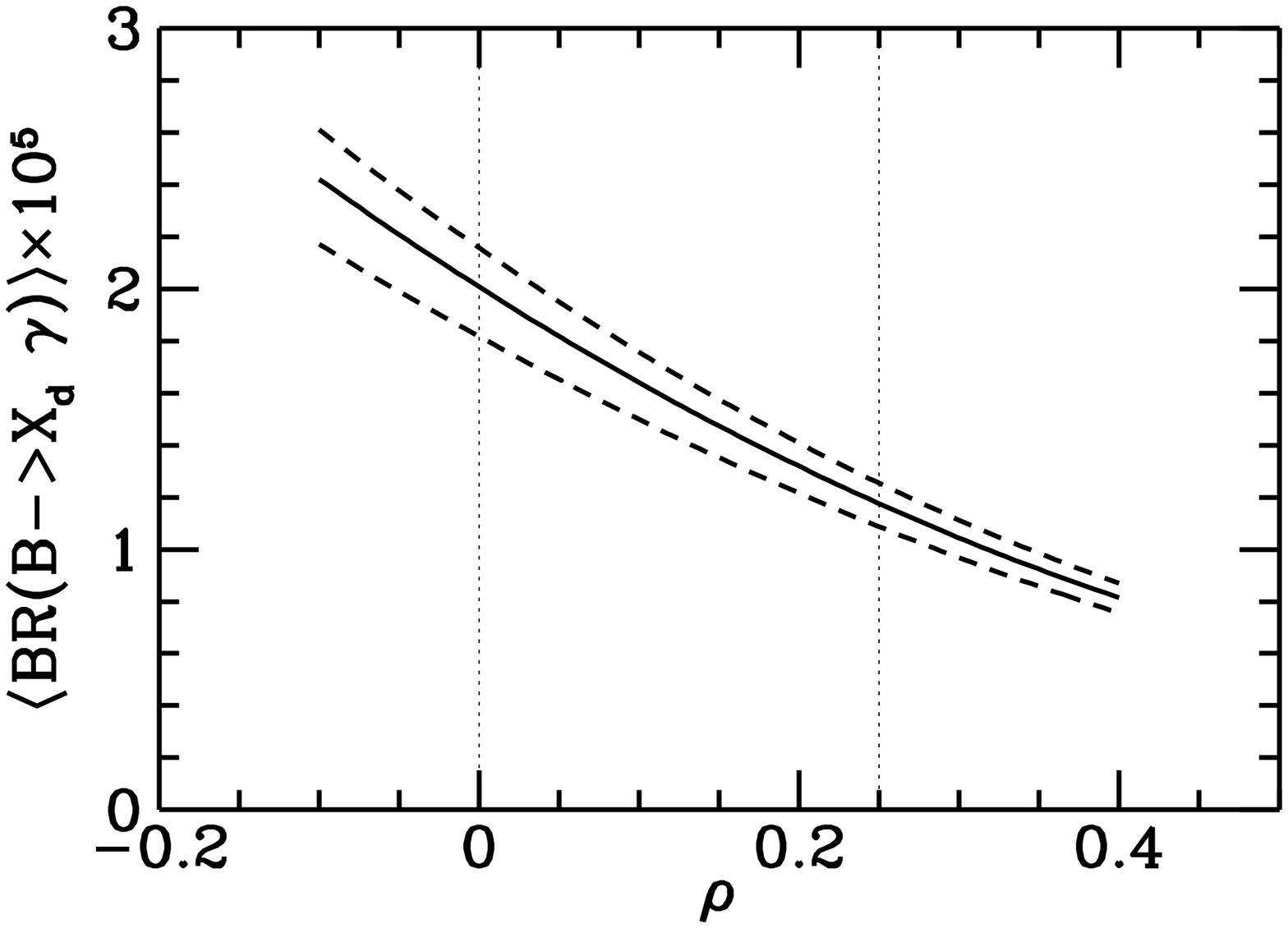}
\end{center}
\caption[]{
$\rho$ dependence of the average branching ratio for $\BGAMAXD$
and $\BbarGAMAXD$ for fixed $\eta=0.33$. 
The solid curve corresponds to $\mu_b=2.5$ GeV and the central values 
of the input parameters.
The upper and lower dashed curves show the theoretical dispersion 
due to the errors in the input parameters
$m_b$, $m_c/m_b$, ${\cal B}_{sl}$, $m_t$, $\alpha_s(m_Z)$,$\alpha_{em}$
and due to the variation of the scale 
$\mu_b \in[2.5\,\mbox{GeV},10.0\,\mbox{GeV}]$.  
The long-distance contribution due to the $u$-quark loop is also included
in estimating the errors (see text).
The vertical lines show the $\pm 1 \sigma$ range for $\rho$ from the
CKM fits \protect\cite{aliapctp97}.
\label{fig4}}
\end{figure}

\begin{figure}
\begin{center}
\epsfxsize=10.5cm
\leavevmode  
\epsfbox[20 187 573 577]{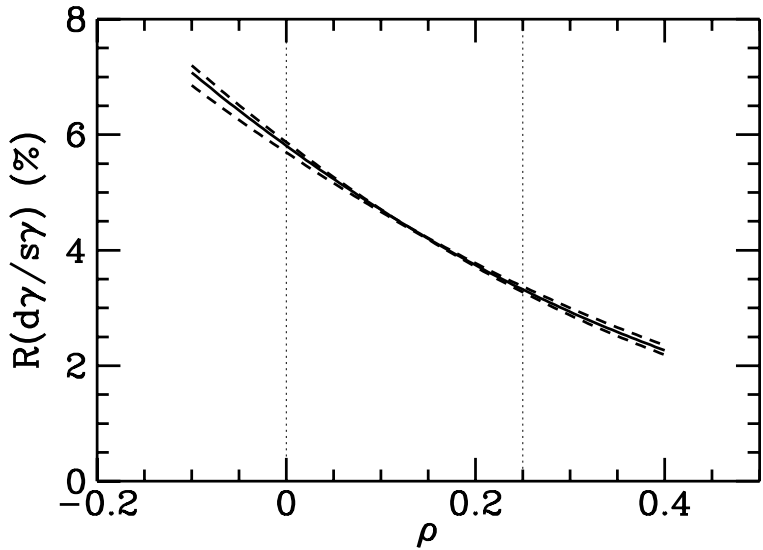}
\end{center}  
\caption[]{
The ratio $R(d\gamma/s\gamma)$ (in \%) as a function of the CKM
parameter $\rho$ for a fixed value of $\eta=0.33$. The bands show the
theoretical uncertainties following from the error estimates
discussed in text. The vertical lines show the $\pm 1 \sigma$ range 
for $\rho$ from the CKM fits \protect\cite{aliapctp97}.
\label{figrdsgam}}
\end{figure}

\begin{figure}
\begin{center}
\epsfxsize=8.5cm
\leavevmode
\epsfbox[20 187 574 712]{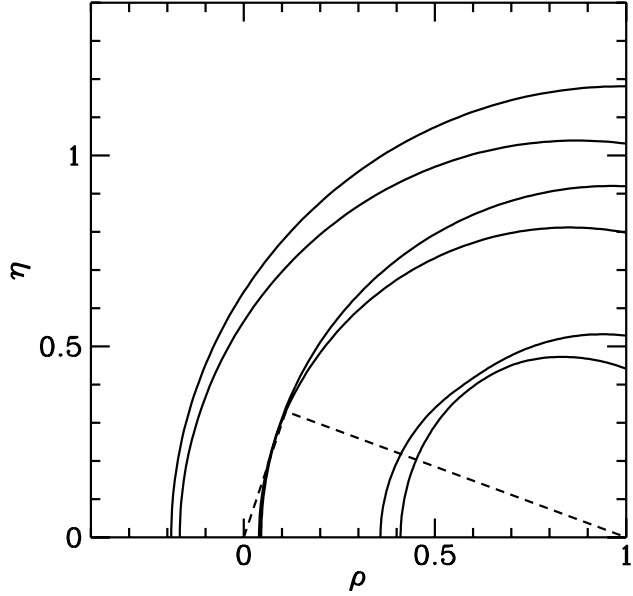}
\end{center}
\caption[]{
Fixed-$R$ contours in the $(\rho,\eta)$ plane, obtained by
varying the input parameters
and the scale $\mu_b \in [2.5\,\mbox{GeV},10.0\,\mbox{GeV}]$.
The three bands shown in the figure
correspond to $R=0.017$ (bottom), $R=0.046$ (middle)
and $R=0.074$ (top). The unitarity triangle corresponding to the 
"best fit" solution from the CKM fits \protect\cite{aliapctp97} is also 
shown.
\label{ratio}} 
\end{figure}

To that end, it is important to estimate the theoretical uncertainties in
the branching ratio, the ratio $R(d\gamma/s\gamma)$, and direct
CP asymmetry in $\BGAMAXD$ and $\BbarGAMAXD$ 
for given values of $\eta$ and $\rho$. 
We estimate these theoretical uncertainties by varying
the scale $\mu_b \in [2.5\,\mbox{GeV},10.0\,\mbox{GeV}]$ and the input
parameters in their respective $\pm 1\sigma$-ranges given earlier.
The procedure adopted is as follows: Individual errors 
$\Delta_i (\BRAV)$, $\Delta_i (R(d\gamma/s\gamma))$ and $\Delta_i(a_{CP})$
are estimated by varying each parameter at a time and
the resulting errors are then added in
quadrature, much the same way as it has been done for estimating the
theoretical uncertainty in the branching ratio
$\BBbarGAMAXS$. As mentioned earlier,
we add an error of $\pm 0.1D_t^{(0)}$ 
in the numerical 
estimate of the function $D_r$ in order to take into account
long-distance effects generated by intermediate $u$-quarks.
The resulting theoretical uncertainty 
on $\BRAV$ from all the sources is shown
in Fig.~\ref{fig4} as
$\pm 1 \sigma$ bands for the central value $\eta=0.33$ as a function of $\rho$.
For a given value of $\rho$ and $\eta$, the theoretical uncertainty is:
$\Delta (\BRAV)/\BRAV = 
\pm (6-10)\%$ on the branching ratio. This is
much smaller than the factor 4
dispersion in $\BRAV$ due to the $\rho$ and $\eta$ dependence, shown
in Fig.~\ref{fig2}.

The uncertainty in the ratio $R(d\gamma/s\gamma)$
is even smaller, since the theoretical errors on
$\langle \BBGAMAXS \rangle$ and $\BRAV$ tend to cancel.
The residual theoretical uncertainty $\Delta R/R$ is correlated with the
value of $\rho$ and $\eta$, which is not difficult to see from the
relation in eq.~(\ref{ratiosd}). For the central value of the CKM 
fits $\eta=0.33$, this is shown in Fig.~\ref{figrdsgam} where we plot
$R(d\gamma/s\gamma)$ as a function of
$\rho$. Interestingly, the theoretical uncertainties almost vanish for
$\rho$ in the proximity of the "best fit" value $\rho =0.11$. The largest
theoretical uncertainty $R(d\gamma/s\gamma)$ in the $95\%$ C.L. allowed 
CKM-domain is for the point ($\rho=0.4$,~$\eta=0.2)$ where 
$\Delta R(d\gamma/s\gamma)/R(d\gamma/s\gamma) =\pm 7\%$, as the
ratio $R(d\gamma/s\gamma)$ is smallest there. This study 
suggests that the impact of the measurement of $R(d\gamma/s\gamma)$ on
the CKM parameters will be largely determined by experimental errors.  

It is interesting   
to see with which theoretical accuracy the Wolfenstein parameters
$\rho$ and $\eta$ get constrained  assuming an ideal
measurement of $R$. To illustrate this, we study 
the fixed-$R$ contours in the $(\rho,\eta)$ plane.
Our procedure
is as follows:
We choose three hypothetical values $R=0.017,0.046,0.074$, emerging from our
NLL analysis. 
For each of these values, we solve eq. (\ref{ratiosd}) for $\eta$. 
Fixing all input parameters, except $\rho$, leads to a curve
in the $(\rho,\eta)$ plane (fixed-$R$ contour).
Varying then the input parameters $\alpha_s(m_z)$, $m_b$, $m_c/m_b$,
$m_t$ and the scale $\mu_b$ (one at a time followed by adding the individual
errors in quadrature), leads to a band in the $(\rho,\eta)$-plane for
each value of $R$. In Fig.~\ref{ratio} these bands are shown for
the values of $R$ indicated above.
 The unitarity triangle corresponding to the "best fit" 
solution $(\rho=0.11,\eta=0.33)$ is also drawn for orientation. One sees
again that the theoretical uncertainties are minimal (practically vanishing)
for the "best fit" solution.
 
\begin{figure}
\begin{center}
\epsfxsize=10.5cm
\leavevmode
\epsfbox[20 187 573 577]{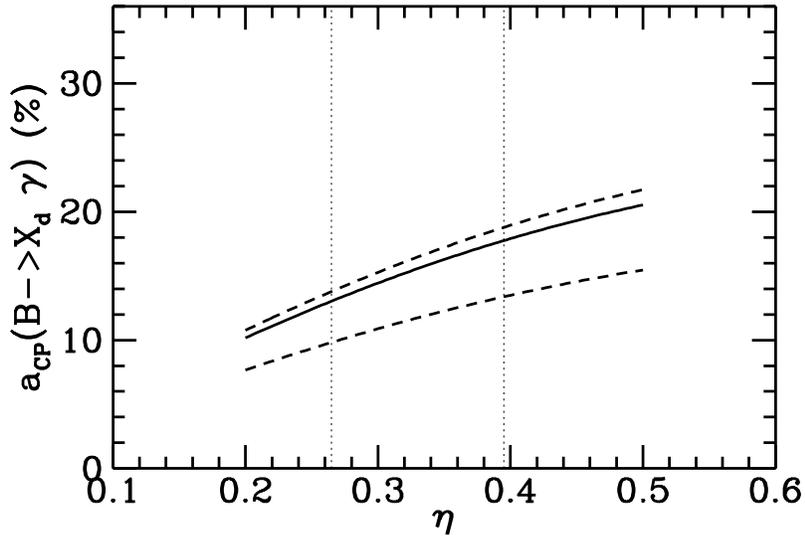}
\end{center}
\caption[]{
$\eta$ dependence of the CP rate asymmetry $a_{CP}(\BGAMAXD)$
for fixed $\rho=0.11$. 
The solid curve corresponds to $\mu_b=2.5$ GeV and the central values 
of the input parameters.
The upper and lower dashed curves show the theoretical dispersion
due to the errors in the parameters
$m_b$, $m_c/m_b$, ${\cal B}_{sl}$, $m_t$, $\alpha_s(m_Z)$, $\alpha_{em}$
and due to the variation of the scale 
$\mu_b \in[2.5\,\mbox{GeV},10.0\,\mbox{GeV}]$.
The vertical lines show the $\pm 1 \sigma$ range for $\eta$ from the
CKM fits \protect\cite{aliapctp97}.
  
\label{fig7}}
\end{figure}

 In Fig.~\ref{fig7}, we show the uncertainty on $a_{CP}$ due to the scale
variation and due to the input parameters
as a function of $\eta$ (with fixed $\rho=0.11$). As mentioned, the
power corrections drop out in the LL approximation. 
For given values of $\rho$ and $\eta$,
we find: $\Delta(a_{CP})/a_{CP}= \pm 17\%$.
Since the
asymmetry $a_{CP}(\BGAMAXD)$ itself varies between $7\%$ and $35\%$ (see
Fig.~\ref{fig3}) in the
presently allowed range of the parameters $\rho$ and $\eta$, the residual
theoretical uncertainty  is
not a serious hindrance in testing the CKM paradigm for CP violation in
these decays. Of course, it will be nice to complete the calculation for
$a_{CP}$ in the NLL approximation, which hopefully will reduce the
theoretical uncertainty on this quantity considerably.\\

{\bf 5. Summary}\\

  To summarize, we have presented theoretical estimates of the branching
ratio $\BRAV$ and the ratio $R(d\gamma/s\gamma)$ 
in the NLL approximation, and $a_{CP}(B \to X_d + \gamma)$ in the
LL approximation in SM, working out also theoretical errors.
Varying the CKM-Wolfenstein parameters $\rho$ and $\eta$ in the range
$-0.1 \leq \rho \leq 0.4$ and $0.2 \leq \eta \leq 0.46$ and taking into
account other parametric dependences stated earlier, our numerical
results can be summarized as follows:
\begin{eqnarray}
\label{summarybrasy}
6.0 \times 10^{-6} &\leq &
\BRAV \leq 2.6 \times 10^{-5}~, \nonumber\\
0.017 &\leq & R(d\gamma/s\gamma) \leq 0.074~,\nonumber\\
0.07 &\leq & a_{CP}(\BGAMAXD) \leq 0.35~.
\end{eqnarray}
The central values of these quantities corresponding to the "best fit"
parameters ($\rho=0.11$,~$\eta=0.33)$ are:
$\BBbarGAMAXD=(1.61 \pm 0.12) \times 10^{-5}$, $R(d\gamma/s\gamma)=0.046$ and
$a_{CP}(\BGAMAXD)= (11.5 - 16.5)\%$, with practically no error on 
$R(d\gamma/s\gamma)$.
This ratio is also otherwise found to be remarkably stable against
variation in the input parameters, with the maximum uncertainty estimated as
$\Delta R(d\gamma/s\gamma)/ R(d\gamma/s\gamma) =\pm 7\%$ for
($\rho=0.4,~\eta=0.2)$.
These quantities are expected to be measurable at the forthcoming high 
luminosity B
facilities. The CP-violating asymmetry 
$a_{CP}(\BGAMAXS)$ in the SM is found to be too small to measure. We
emphasize the need to complete the NLL-improved calculation for
$a_{CP}(\BGAMAXD)$.\\

 We would like to thank Francesca Borzumati and Daniel Wyler for 
helpful discussion.
This work was partially supported by INTAS under the Contract INTAS-96-155.

\newpage
\end{document}